# Topological GCN for Improving Detection of Hip Landmarks from B-Mode Ultrasound Images

Tianxiang Huang[1], Jing Shi[2], Ge Jin[1,3], Juncheng Li[1], Jun Wang[1], Jun Du[2(✉)], and Jun Shi[1(✉)]

[1] School of Communication and Information Engineering, Shanghai University, Shanghai 200444, China
`junshi@shu.edu.cn`
[2] Diagnostic Imaging Center, Shanghai Children's Medical Center, School of Medicine, Shanghai Jiao Tong University, Shanghai, 200127, China
`dujun@scmc.com.cn`
[3] School of Communication and Information Engineering, Jangsu Open University, Jiangsu 214257, China

**Abstract.** The B-mode ultrasound based computer-aided diagnosis (CAD) has demonstrated its effectiveness for diagnosis of Developmental Dysplasia of the Hip (DDH) in infants. However, due to effect of speckle noise in ultrasound images, it is still a challenge task to accurately detect hip landmarks. In this work, we propose a novel hip landmark detection model by integrating the Topological GCN (TGCN) with an Improved Conformer (TGCN-ICF) into a unified framework to improve detection performance. The TGCN-ICF includes two subnetworks: an Improved Conformer (ICF) subnetwork to generate heatmaps and a TGCN subnetwork to additionally refine landmark detection. This TGCN can effectively improve detection accuracy with the guidance of class labels. Moreover, a Mutual Modulation Fusion (MMF) module is developed for deeply exchanging and fusing the features extracted from the U-Net and Transformer branches in ICF. The experimental results on the real DDH dataset demonstrate that the proposed TGCN-ICF outperforms all the compared algorithms.

## 1 Introduction

Developmental dysplasia of the hip (DDH) is one of the most common orthopedic disorders in infants, which may lead to acetabular dysplasia, hip instability, and hip dislocation [1]. Accurate diagnosis of DDH in the early stagy is crucial for the following treatment [2]. In clinical practice, B-mode Ultrasound (BUS) imaging is commonly used for diagnosis of DDH in infants within 6 months [3]. The Graf's method is commonly used for diagnosing DDH by measuring the $\alpha$ and $\beta$ angles (Normal if $\alpha > 60°$ and $\beta < 77°$) as shown in Fig. 1(a) [4]. However, this method is susceptible to the subjective expertise of sonologists.

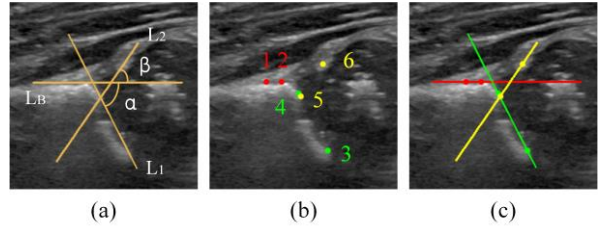

**Fig. 1.** Illustration of hip BUS images.

In recent years, deep learning (DL) has gained its reputation in the field of BUS-based computer-aided diagnosis (CAD) for DDH [5-8]. Since the angle measurement can be simply determined by some critical hip landmarks as shown in Fig. 1(b) and (c), some pioneering works have explored the feasibility of CAD based on hip landmark detection [9, 10]. However, accurate detection of key hip landmarks is still a challenging task due to the effect of speckle noise in ultrasound images [11].

It is worth noting that the hip landmarks are located across multiple areas of the BUS images (as shown in Fig. 1(b)), and therefore, it is necessary to capture both the local and global information to improve detection accuracy. Since the Convolutional Neural Network (CNN) mainly focuses on extracting local features [12], and the Transformer architecture can well learn global representations [13], the hybrid models by combining CNN and Transformer then indicate their superior performance for the point detection task [14, 15]. As a classical hybrid model, Conformer has shown its effectiveness in many computer vision tasks [16], and has the feasibility for hip landmark detection. However, the feature fusion strategy in the Conformer is very simple, and cannot fully fuse the local and global features extracted from the CNN and Transformer branches, which will affect the detection performance to a certain extend.

On the other hand, as shown in Fig. 1(b), the key landmarks have their inherent spatial relations in the hip BUS images. For example, the red landmarks (landmark 1 and 2) are collinear, which serve as the key points to form the base line ($L_B$ in Fig. 1(a)) according to the Graf's method [4]. These special spatial relations between different landmarks can provide important spatial topology knowledge to help enhance the detection accuracy. However, existing hip landmark detection algorithms do not pay close attention to this important prior information, and it is also difficult to model and utilize it. Since the Graph Convolutional Network (GCN) can effectively integrate the topology information into a graph [17, 18], it provides a feasible approach to capture the spatial relations of landmarks for further improving detection performance.

In this work, we propose a novel hip landmark detection model by integrating the Topological GCN (TGCN) with an Improved Conformer (TGCN-ICF) into a unified framework to improve detection performance. The TGCN-ICF includes two subnetworks: an Improved Conformer (ICF) subnetwork to generate the related heatmaps and a TGCN subnetwork to refine landmark detection. This TGCN can effectively improve detection accuracy with the guidance of class labels. Moreover, a Mutual Modulation Fusion (MMF) is developed for deeply exchanging and fusing features extracted from the U-Net and Transformer branches in ICF. The experimental results on a real DDH BUS dataset indicate the effectiveness of the proposed TGCN-ICF.

The main contributions of this work are summarized as follows:
1) A novel unified framework, named TGCN-ICF is proposed for hip landmark detection from BUS images. Different from the conventional landmark detection models that directly detect points based on the generated heatmaps, the additional TGCN subnetwork in TGCN-ICF further refines the heatmaps generated from the ICF subnetwork by learning the spatial topological relations among landmarks with the guidance of class labels, so as to effectively improve detection accuracy.
2) A new MMF model is developed in the ICF subnetwork to fully exchange and fuse the local and global features. In MMF, the local features and global features are optimized by referring to each other, allowing each branch to learn features highly related to itself but missing. In other words, MMF can adaptively suppress the difference patterns of local and global features to achieve fully feature fusion.

## 2   Method

The overall framework of the proposed TGCN-ICF is illustrated in Fig. 2. The TGCN-ICF consists of an ICF subnetwork and a TGCN subnetwork with the following training pipeline.
1) A hip BUS image and the corresponding patches are first fed into the ICF subnetwork to generate heatmaps.
2) The generated heatmaps are then fed into the TGCN subnetwork for further refinement with the guidance of class labels.

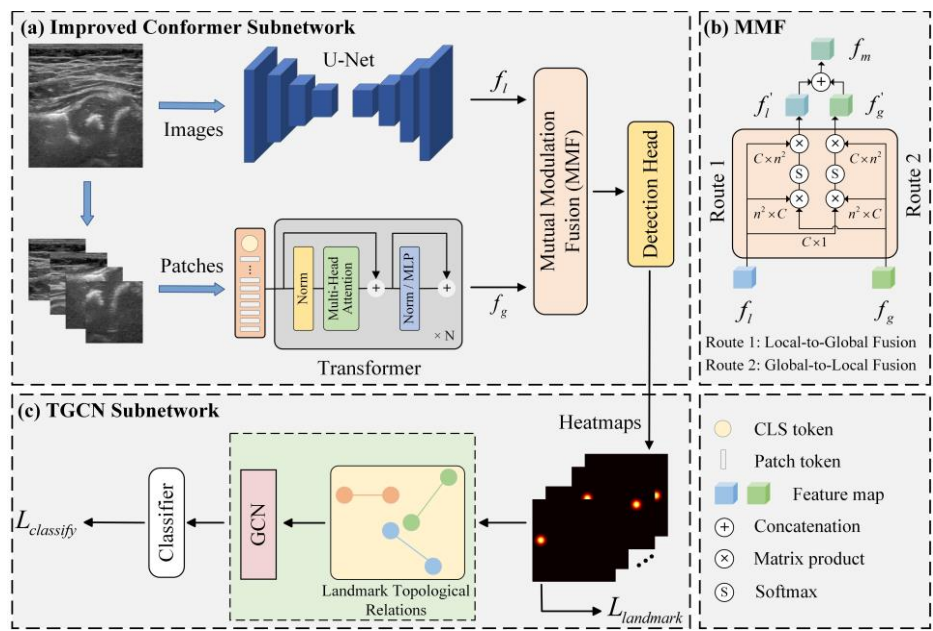

**Fig. 2.** Overview of the proposed Topological GCN with Improved Conformer (TGCN-ICF).

## 2.1 Improved Conformer Subnetwork

In the ICF subnetwork, since U-Net is a commonly used encoder-decoder architecture for landmark detection, we replace the conventional CNN branch in the original Conformer with U-Net. Thus, the U-Net and Transformer branches can effectively capture both the local and global information in hip BUS images. Meanwhile, inspired by [19] and [20], a MMF module is developed to deeply exchange and fuse the features extracted from these two branches. The MMF module can adaptively update and optimize each branch's information by referring to another branch, achieving highly effective fusion of local and global features.

As shown in Fig. 2 (b), given two feature maps $f_l \in \mathbb{R}^{h \times w \times c}$ and $f_g \in \mathbb{R}^{h \times w \times c}$ that are extracted from the U-Net and Transformer, we specifically design two synchronous fusion routes: **Local-to-Global Fusion** and **Global-to-Local Fusion**.

**Local-to-Global Fusion.** In this fusion route, the $f_l$ is updated by $f_g$ in the pixel level. Specifically, a filter $F^{lg}(\cdot)$ is learned to update the local neighbor pixels (denoted as $L^{n^2}_{(i,j)}$) in an $n \times n$ neighborhood by the corresponding center pixel $G_{(i,j)}$ in $f_g$. The filter weight is defined as follows:

$$\boldsymbol{\omega}^0_{(i,j)} = softmax(\sum_c(G_{(i,j)} \otimes L^{n^2}_{(i,j)})) \tag{1}$$

where $softmax(\cdot)$ represents the normalized exponential function, $\sum_c(\cdot)$ denotes the summation along the channel dimension, and $\otimes$ is the matrix product. Therefore, the updated neighbor pixels can be calculated by:

$$L^{n^2}_{(i,j)}{}' = F^{lg}\left[L^{n^2}_{(i,j)}\right] = \sum_{n^2}(L^{n^2}_{(i,j)} \otimes \boldsymbol{\omega}^0_{(i,j)}) \tag{2}$$

where $L^{n^2}_{(i,j)}{}'$ denotes the updated local pixels in the $n \times n$ neighborhood, and $\sum_{n^2}(\cdot)$ represents the summation along the neighborhood spatial dimension. In this way, all pixels in $f_l$ will be updated by targeting the counterpart pixels in $f_g$, and then we can obtain the fused local-to-global information $f_l'$.

**Global-to-Local Fusion.** Similarly, the $f_g$ is updated by $f_l$ in the pixel level. Specifically, a filter $F^{gl}(\cdot)$ is learned to update the global neighbor pixels (denoted as $G^{n^2}_{(i,j)}$) by the corresponding center pixel $L_{(i,j)}$ in $f_l$. The weight of $F^{gl}(\cdot)$ is calculated as:

$$\boldsymbol{\omega}^1_{(i,j)} = softmax(\sum_c(L_{(i,j)} \otimes G^{n^2}_{(i,j)})) \tag{3}$$

Thus, the updated $G^{n^2}_{(i,j)}{}'$ can be obtained by:

$$G^{n^2}_{(i,j)}{}' = F^{gl}\left[G^{n^2}_{(i,j)}\right] = \sum_{n^2}(G^{n^2}_{(i,j)} \otimes \boldsymbol{\omega}^1_{(i,j)}) \tag{4}$$

Therefore, we can get the fused information $f_g{'}$ by updating all pixels in $f_g$. Finally, the fused features $f_l{'}$ and $f_g{'}$ will further be added to obtain the final fused features $f_m$:

$$f_m = f_l{'} \oplus f_g{'} \tag{5}$$

where $\oplus$ represents the concatenation along the channel dimension. In this way, the local and global features can be fully exchanged and fused for subsequent detection.

## 2.2 TGCN Subnetwork

The topological interaction of different landmarks is important and reliable prior information for improving detection performance. However, existing landmark detection-based algorithms ignore the topological information hidden in hip BUS images. Since we can get the class label (normal or abnormal) of each BUS image, we then propose a TGCN subnetwork to effectively learn topology-aware graph representations with the guidance of class labels. It can further refine the generated heatmaps to improve detection accuracy, because the label information can further implicitly provide an additional constraint to correct the detected landmarks.

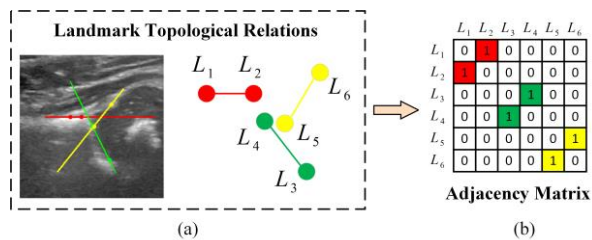

**Fig. 3.** The principle of the adjacency matrix construction.

**Landmark Topological Relations.** As shown in Fig. 3(a), we model three groups of topological relationships from all the hip landmarks inspired by the Graf's method [4]. That is, three critical lines of the related structures are formed by $L_1$ and $L_2$, $L_3$ and $L_4$, $L_5$ and $L_6$, respectively, in Fig. 1(c). In order to make full use of this valuable topological information, we construct a graph for graph representation learning.

A graph is denoted as $G = (V, E)$, where $V$ and $E$ represent the nodes and a set of edges in the graph, respectively. Since each heatmap generated by the ICF subnetwork represents a corresponding landmark, we take each heatmap as a node. Therefore, a graph can be denoted as a feature matrix $G_f \in \mathbb{R}^{k \times d}$, which has $k$ nodes, and each node has a $d$-dimensional feature vector. In this work, we set $k = 6$, since the goal of this task is to extract 6 landmarks, and $d = h \times w$ represents the size of each heatmap.

**Adjacency Matrix Construction.** As shown in Fig. 3 (b), we construct the adjacency matrix $A_{i,j} \in \mathbb{R}^{k \times k}$ based on the above mentioned three groups of topological relations.

Specifically, the adjacency matrix can be denoted as follows:

$$A_{i,j} = \begin{cases} 1, & (v_i, v_j) \in E \\ 0, & otherwise \end{cases} \tag{6}$$

where $(v_i, v_j)$ denotes an edge. For example, since $L_1$ and $L_2$ are collinear, we define $A_{1,2} = A_{2,1} = 1$. Thus, six edges of the graph can be constructed from the three groups of hip landmark topological relations. In this way, the valuable topology information can be encoded into an adjacency matrix for further learning graph representations.

After obtaining $G_f \in \mathbb{R}^{k \times d}$ and $A_{i,j} \in \mathbb{R}^{k \times k}$, they are sent into a muti-layer GCN [21]. The operation can be denoted as follows:

$$G_f{'} = \sigma(\widetilde{D}^{-\frac{1}{2}} \tilde{A} \widetilde{D}^{-\frac{1}{2}} G_f W^{(l)}) \tag{7}$$

where $G_f{'} \in \mathbb{R}^{k \times d}$, $\sigma(\cdot)$ denotes an activation function, $\widetilde{D}$ is the degree matrix of $\tilde{A}$, $\tilde{A} = A_{i,j} + I_N$, $I_N$ is the identify matrix, and $W^{(l)}$ is a trainable weight matrix. The graph representations are then fed into the liner projections to obtain the final output:

$$y_{GCN} = (G_f{'} W_0) W_1 \tag{8}$$

where $W_0 \in \mathbb{R}^{d_m \times d}$ and $W_1 \in \mathbb{R}^{d_c \times d_m}$ are the weight matrix of linear projection, $d_m$ and $d_c$ represent the middle dimensionality and final classes number, respectively.

### 2.3 Loss Function

As shown in Fig. 2, TGCN-ICF is trained by two loss functions that consists of the $L_{landmark}$ and $L_{classify}$. Specifically, the former is calculated by the Mean Square Error (MSE) loss between the ground truth heatmaps and predicted heatmaps, and the latter is evaluated by the Binary Cross Entropy (BCE) loss between ground truth labels and predicted classes. The total loss function $L$ is calculated as:

$$L = L_{landmark} + \lambda * L_{classify} \tag{9}$$

where $\lambda$ is a hyperparameter used to adjust the proportion of the two losses.

## 3 Experiments and Results

### 3.1 Datasets

To evaluate the effectiveness of TGCN-ICF, we conducted experiments on a real world DDH dataset. This dataset includes 500 hip ultrasound images (458 normal subjects and 42 abnormal subjects) from 294 infants, which were captured by the LOGIQ E9 ultrasound scanner (GE HealthCare, Milwaukee, WI). Notably, all landmarks were marked by experienced sonologists.

### 3.2 Experiment Setup

To evaluate the performance of the proposed TGCN-ICF, we compared it with the following algorithms:

1) U-Net [22]: The classical U-Net model was applied for landmark detection.
2) DM-ResNet [9]: It was specially proposed model for hip landmark detection task, which adopted a simple ResNet as the backbone and presented a novel dependency mining module to enhance feature representation.
3) Conformer [16]: It was the original Conformer model but with the U-Net instead of the CNN branch, which was compared as a baseline.
4) FAT-Net [23]: This model was a representative dual-branch network that utilized the CNNs and Transformer as a dual encoder with three feature adaptation modules.
5) DA-TransUNet [24]: This model was a newly proposed U-shape architecture, which utilized Transformers and dual attention blocks to integrate both global and local features together with the image-specific positional and channel features.

We also conducted an ablation experiment to further evaluate the effectiveness of the proposed MMF and TGCN:

1) TGCN-ICF without MMF (TGCN-ICF w/o MMF): This variant used the conventional concatenation strategy instead of MMF to fuse the features of U-Net and Transformer branches in TGCN-ICF.
2) TGCN-ICF without TGCN (TGCN-ICF w/o TGCN): This variant removed the proposed TGCN subnetwork, and then directly applied the improved Conformer subnetwork for detecting hip landmarks.

We performed the five-fold cross-validation to evaluate the performance of all the algorithms. The commonly used mean radial error (MRE) and successful detection rate (SDR) were adopted as the evaluation indices [9]. All the results were presented in the format of mean ± SD (standard deviation).

### 3.3 Implementation Details

In our implementations, the Random Horizontal Flip was utilized as a data augmentation operation. The model was trained by an Adam optimizer with an initial leaning rate of 1e-4. Moreover, the TGCN-ICF was trained for 100 epochs with a batch size of 2. All the algorithms were implemented by PyTorch with a GTX 3090 GPU.

### 3.4 Experimental Results

Fig. 4 shows the visualization results of hip landmark detection by different algorithms. The red dots and green dots represent the positions of ground truth landmarks and predicted landmarks, respectively. It can be found that the proposed TGCN-ICF achieves the best detection accuracy compared to other algorithms. Moreover, when removing the MMF or TGCN, the predicted landmarks exhibit a deviation from the ground truth landmarks compared with the proposed TGCN-ICF, which suggests the effectiveness of proposed MMF and TGCN, respectively.

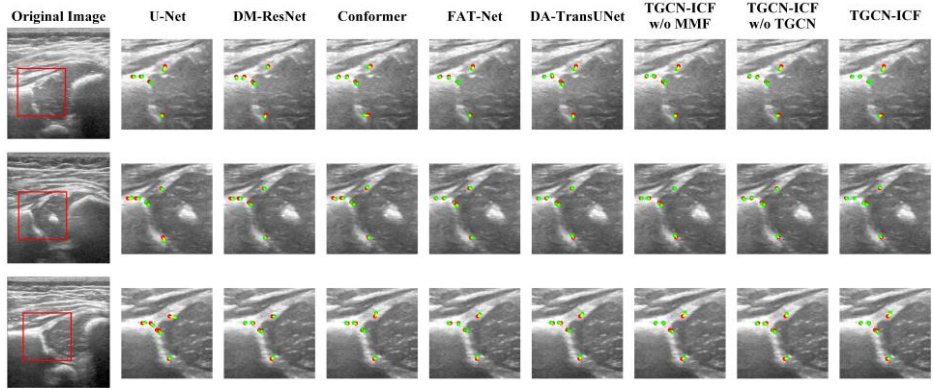

**Fig. 4.** Visual comparison of different landmark detection algorithms.

Table 1 gives the quantitative results of different algorithms. It can be observed that the proposed TGCN-ICF outperforms all the compared algorithms with the best MRE of 0.4364±0.0388mm and three SDRs of 72.33±1.19 (0.5mm), 94.73±1.23% (1.0mm), and 98.47±1.42% (1.5mm), respectively. Compared to others, TGCN-ICF decreases at least 0.0087mm (approximately 1.95%) on MRE, and it also improves at least 1.83%, 0.53%, and 0.17% on the three SDRs, respectively. These superior results demonstrate the effectiveness of TGCN-ICF for detecting hip landmarks.

**Table 1.** Quantitative results of different algorithms for hip landmark detection.

| Method | MRE (mm) ↓ | SDR (%) ↑ | | |
| --- | --- | --- | --- | --- |
| | | 0.5mm | 1.0mm | 1.5mm |
| U-Net [22] | 0.5147±0.0349 | 64.80±2.81 | 92.17±0.72 | 97.40±1.37 |
| DM-ResNet [9] | 0.5057±0.0468 | 66.43±2.56 | 92.27±1.14 | 97.37±1.18 |
| Conformer [16] | 0.4657±0.0314 | 67.93±2.81 | 93.53±0.72 | 98.30±1.37 |
| FAT-Net [23] | 0.4451±0.0446 | 70.50±1.80 | 94.20±1.12 | 98.30±1.17 |
| DA-TransUNet [24] | 0.4537±0.0376 | 70.26±0.62 | 94.10±1.08 | 97.80±1.39 |
| **TGCN-ICF (Ours)** | **0.4364±0.0388** | **72.33±1.19** | **94.73±1.23** | **98.47±1.42** |

Table 2 shows the quantitative results on the ablation study. When employing the traditional concatenation strategy for fusion, the TGCN-ICF w/o MMF exhibits an increase of 0.0079mm (about 1.78%) on MRE, and reductions of 0.96%, 0.70%, and 0.34% on SDRs, compared with the TGCN-ICF. It suggests the effectiveness of MMF module in exchanging and fusing features from U-Net and Transformer branches. In comparison to TGCN-ICF, the variant TGCN-ICF w/o TGCN increases 0.0101mm (approximately 2.26%) on MRE, and reduces 1.40%, 0.56%, and 0.10% on three SDRs, respectively. These results prove the importance of the TGCN to learn topology graph representation for enhancing the hip landmark detection performance. Moreover, both TGCN-ICF w/o MMF and TGCN-ICF w/o TGCN outperform the Conformer baseline.

These superior results once again indicate the effectiveness of the proposed MMF module and TGCN subnetwork.

Table 2. Quantitative results of ablation study for hip landmark detection.

| Method | MRE (mm) ↓ | SDR (%) ↑ | | |
|---|---|---|---|---|
| | | 0.5mm | 1.0mm | 1.5mm |
| Conformer (Baseline) | 0.4657±0.0314 | 67.93±2.81 | 93.53±0.72 | 98.30±1.37 |
| TGCN-ICF w/o MMF | 0.4443±0.0392 | 71.37±1.28 | 94.03±1.23 | 98.13±1.24 |
| TGCN-ICF w/o TGCN | 0.4465±0.0431 | 70.93±1.81 | 94.17±1.04 | 98.37±1.47 |
| **TGCN-ICF (Ours)** | **0.4364±0.0388** | **72.33±1.19** | **94.73±1.23** | **98.47±1.42** |

## 4    Conclusion

In conclusion, we propose a novel TGCN-ICF for landmark detection within hip BUS images. The TGCN-ICF can learn valuable topology graph representation with the guidance of class label to improve landmark detection. Moreover, a new MMF module is developed for effectively exchanging features between two branches in ICF subnetwork. The experimental results show the effectiveness of the proposed TGCN-ICF, suggesting its potential application for the CAD of DDH.

**Acknowledgments.** This work is supported by the National Natural Science Foundation of China (62271298), the 111 Project (D20031), the Fund of the Cyrus Tang Foundation, and the Fund of the Education Development Foundation of Shanghai Jiao Tong University.

**Disclosure of Interests.** The authors have no competing interests to declare that are relevant to the content of this article.

## References

1. Sewell, M.D.: Rosendahl, K., Eastwood, D.M.: Developmental dysplasia of the hip. Bmj. **339** (2009)
2. Sioutis, S., et al.: Developmental dysplasia of the hip: a review. J. Long-Term Eff. Med. Implants. **32**(3), 39-56 (2022)
3. Zhang, D., et al.: Multi-frequency therapeutic ultrasound: A review. Ultrason. Sonochem. **100** (2023)
4. Graf, R.: Fundamentals of sonographic diagnosis of infant hip dysplasia. J. Pediatr Orthop. **4**(6), 735-740 (1984)
5. Golan, D., et al.: Fully automating Graf's method for DDH diagnosis using deep convolutional neural networks. Deep learn. Data Label. Med. Appl. **10008**, 130-141 (2016)
6. Lee, SW., et al.: Accuracy of new deep learning model-based segmentation and key-point multi-detection method for ultrasonographic developmental dysplasia of the hip (DDH) screening. Diagnostics, **11**(7), 1174 (2021)

7. Stamper, A., et al.: Infant hip screening using multi-class ultrasound scan segmentation. In: ISBI, pp. 1-4 (2023)
8. Liu, R., et al.: NHBS-Net: A feature fusion attention network for ultrasound neonatal hip bone segmentation. IEEE Trans. Med. Imag. **40**(12), 3446-3458 (2021)
9. Xu, J., et al.: Hip landmark detection with dependency mining in ultrasound image. IEEE Trans. Med. Imag. **40**(12), 3762-3774 (2021)
10. Chen, YP., et al.: Automatic and Human Level Graf's Type Identification for Detecting Developmental Dysplasia of the Hip. Biomed. J. 100614 (2023)
11. Pradeep, S., Nirmaladevi, P.: A review on speckle noise reduction techniques in ultrasound medical images based on spatial domain, transform domain and CNN methods. IOP Conf. Ser.: Mater. Sci. Eng. **1055**(1), 012116 (2021)
12. Kshatri, SS., Singh, D.: Convolutional neural network in medical image analysis: a review. Arch. Comput. Methods Eng. **30**(4), 2793-2810 (2023)
13. Shamshad, F., et al.: Transformers in medical imaging: A survey. Med. Image Anal. **88**, 102802 (2023)
14. Viriyasaranon, T., Ma, S., Choi, JH.: Anatomical Landmark Detection Using a Multiresolution Learning Approach with a Hybrid Transformer-CNN Model. In: Greenspan, H., et al. (eds.) MICCAI 2023. LNCS, vol. 14225, pp. 433-443, Springer, Cham (2023). https://doi.org/10.1007/978-3-031-43987-2_42
15. Wu, F., et al.: Multi-scale Hybrid Transformer Network with Grouped Convolutional Embedding for Automatic Cephalometric Landmark Detection. In: Hu, S.-M., et al. (eds.) CAD/Graphics 2023. LNCS, vol. 14250, pp. 250-265, Springer, Cham (2024). https://doi.org/10.1007/978-981-99-9666-7_17
16. Peng, Z., et al.: Conformer: Local features coupling global representations for visual recognition. In: WACV, pp. 367-376 (2021)
17. Zhang, T., et al.: Topological structure and global features enhanced graph reasoning model for non-small cell lung cancer segmentation from CT. Phys. Med. Biol. **68**(2), 025007 (2023)
18. Wang, Y., et al.: HandGCNFormer: A Novel Topology-Aware Transformer Network for 3D Hand Pose Estimation. In: WACV, pp. 5664-5673 (2023)
19. Dong, X., et al.: Learning mutual modulation for self-supervised cross-modal super-resolution. In: Avidan, S., et al. (eds.) ECCV 2022. LNCS, vol. 13679, pp. 1-18. Springer, Cham (2022). https://doi.org/10.1007/978-3-031-19800-7_1
20. Qiao, X., et al.: Self-supervised depth super-resolution with contrastive multiview pre-training. Neural Netw. **168**, 223-236 (2023)
21. Kipf, TN., Welling, M.: Semi-supervised classification with graph convolutional networks. In: ICLR, pp. 1-14 (2017)
22. Ronneberger, O., Fischer, P., Brox, T.: U-net: Convolutional networks for biomedical image segmentation. In: Navab, N., et al. (eds.) MICCAI 2015, LNCS, vol. 9351, pp. 234–241. Springer, Cham (2015). https://doi.org/10.1007/978-3-319-24574-4_28
23. Wu, H., et al.: FAT-Net: Feature adaptive transformers for automated skin lesion segmentation. Med. Image Anal. **76**, 102327 (2022)
24. Sun, G., et al.: DA-TransUNet: Integrating Spatial and Channel Dual Attention with Transformer U-Net for Medical Image Segmentation. arXiv preprint arXiv:2310.12570 (2023)